\documentclass[11pt]{article}
\usepackage{moriond,epsfig}
\usepackage{graphicx}
\usepackage{fancyhdr}
\usepackage{amsmath,amssymb}
\usepackage[usenames,dvipsnames]{color}
\usepackage{aas_macros}

\def\be{\begin{equation}}
\def\ee{\end{equation}}
\def\bea{\begin{eqnarray}}
\def\eea{\end{eqnarray}}

\begin{document}
\vspace*{4cm}
\title{Cosmic ray constraints on singlino-like dark matter candidates}

\author{Timur DELAHAYE \footnote{presenting author}, David CERDE\~NO \& Julien LAVALLE}

\address{Instituto de F\'isica Te\'orica UAM/CSIC\\
Universidad Aut\'onoma de Madrid
Cantoblanco, 28049 Madrid, Spain}

\maketitle\abstracts{
Recent results from direct detection experiments (Dama, CoGeNT), though subject to debate, seem to point toward a low mass (few GeV) dark matter (DM) particle. However, low mass DM candidates are not easily achieved in the MSSM nor NMSSM. As shown by some authors, singlet extensions of the MSSM can lead to GeV mass neutralinos and satisfy relic abundance constraints. We propose here to extract indirect detection constraints on these models in a generic way from cosmic-ray anti-proton measurements (PAMELA data)}

Recent results from dark matter direct detection experiments DAMA~\cite{Bernabei:2010mq} and CoGeNT~\cite{2011PhRvL.106m1301A} have risen a lot of interest for light dark matter (masses between 3 an 20 GeV, roughly). Though these results are in tension with those of the Xenon100~\cite{2010PhRvL.105m1302A} experiment (in the case of a pure spin-independent scattering), the light dark matter hypothesis is an interesting one to explore.

As it has been shown by Julien Lavalle~\cite{2010PhRvD..82h1302L}, interpreting the results from DAMA~\cite{Bernabei:2010mq} and CoGeNT~\cite{2011PhRvL.106m1301A} in terms of dark matter may be in tension with anti-proton cosmic rays constraints. Indeed a dark matter candidate that has such a large coupling to quarks (as suggested by the CoGeNT results) may imply a strong annihilation into quark pairs which, may produce more cosmic ray anti-protons in the Galactic halo than what has been observed by PAMELA~\cite{2010PhRvL.105l1101A}. It has been shown~\cite{2011PhRvD..83a5001F} that some part of phase of the MSMM may survive the cosmic-ray constraints, however only if the astrophysical parameters are very favourable.

Following the idea of other works \cite{2011PhLB..695..169K,2011arXiv1104.0679K,2010arXiv1009.2555G,2011PhRvL.106l1805D}, we have investigated the possibility that rather than having a dark matter particle annihilating into lepton pairs, it would annihilate into scalar and pseudo-scalar particles which, in turn would decay into Standard Model fermions. There still would be a cosmic anti-proton production, however, which a much less sharp spectrum, and, in some cases, compatible with the PAMELA measurements. The Next-to-Minimal SuperSymetric Model (NMSSM) may have such a particle content (in the case where the dark matter particle is mainly singlino it can be made very light and annihilate into the scalar and the pseudo-scalar Higgses). However NMSSM is not a requirement and any particle physics model with a similar particle content would give comparable results.

\section{Particle content and cross sections}

\begin{figure}[!h]
\begin{center}
\includegraphics[width=0.48\textwidth]{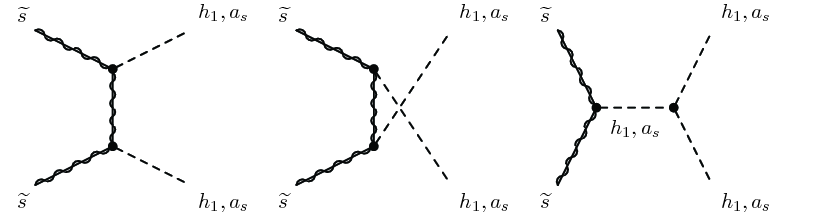}
\end{center}
\caption{Dark matter annihilation channels considered.
\label{fig:fdiag}}
\end{figure}

The phenomenology of singlino-like dark matter scenarios is mostly set by the couplings between the neutralino, a Majorana fermion, and the light 
scalar and pseudo-scalar particles of the Higgs sector. Denoting
$\chi$ the neutralino field and $\phi_i$ the scalar or pseudo-scalar fields,
the effective Lagrangian that we consider reads:

\begin{equation}
{\cal L}_{\rm eff} = 
-\frac{1}{2}\sum_i\chi\,{\cal C}_{\chi i}\,\chi\,\phi_i
-\frac{1}{2}\sum_{i\leq j\leq k}\lambda_{ijk}\,\phi_i\,\phi_j\,\phi_k\;, \nonumber
\label{eq:lag}
\end{equation}

where we $\lambda_{i,j,k}$ is a dimensional coupling taken either fully real or 
fully imaginary, and where
\begin{eqnarray}
\label{eq:coupl}
{\cal C}_{\chi i} &\equiv& c_{\chi\chi i} +  \tilde{c}_{\chi\chi i} \gamma_5  \nonumber\\
\bar{{\cal C}}_{\chi i} &\equiv& c_{\chi\chi i} -  \tilde{c}_{\chi\chi i} \gamma_5  \nonumber
\end{eqnarray}
features both the couplings of the scalar and pseudo-scalar fields $\phi_i$ to 
the dark matter fermionic field $\chi$. With this parametrisation, considering the diagrams of Figure~\ref{fig:fdiag}, the annihilation cross-section can be computed analytically (see details in\cite{mainpaper}). 

All we have to do is hence to restrict the parameter space of masses and coupling that is compatible with the CoGeNT region and collider constraints.

\section{Relic Density}

In the standard $\Lambda$-Cold Dark Matter model ($\Lambda$CDM), the dark matter we consider here is thermally produced during the first times of the universe. Initially in chemical and thermal equilibrium with the plasma, its abundance obeys standard quantum statistical equilibrium.  When the temperature drops below their mass, dark matter particles still annihilate and experience Boltzmann suppression until the expansion rate of the universe, controlled by the effective degrees of freedom $g_\star$, gets larger than the annihilation rate (see e.g. \cite{1991NuPhB.360..145G}). At that time ($T\sim \frac{m_\chi}{20}$), dark matter freezes out, and its relic comoving density is fixed. The present abundance is $\Omega h^2 \propto \frac{1}{g_\star^{1/2} \langle \sigma v \rangle }$. It is important to note that when considering low mass dark matter, the decoupling from the thermal bath occurs at a temperature close to the one of the QCD phase transition which strongly affects the value of $g_\star$.


\begin{figure}[!h]
\begin{center}
\includegraphics[width=0.48\textwidth]{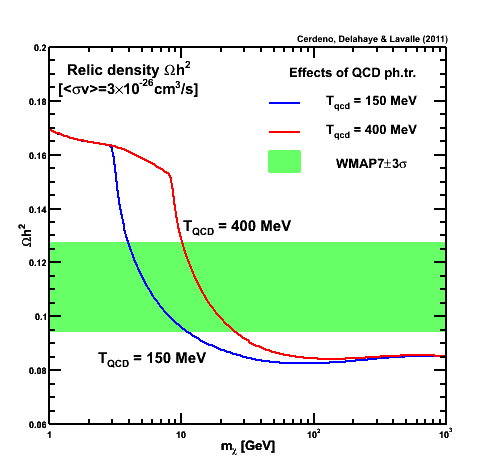}
\end{center}
\caption{Dark matter relic density as a function of dark matter particle mass. The green band corresponds to WMAP7 data. The red and blue lines are the results when considering a first order QCD phase transition happening at 400 MeV and 150 MeV respectively.
}\label{fig:omega}
\end{figure}

As one can see from Figure~\ref{fig:omega}, for a dark matter particle of 10~GeV, the relic density may vary by 60\% depending on the QCD phase transition, which is much more than the observational uncertainty. This consideration is not new at all, however, as it is more usual to consider higher dark matter masses, it is worth recalling.

Because of the high accuracy of WMAP7 data \cite{2011ApJS..192...18K}, the relic density is extremely constraining and as one can see from Figure~\ref{fig:exclus}, it reduces considerably the number of possible values of the couplings of our model. With this reduced parameter space it is now possible to test the cosmic anti-proton flux.

\section{The pbar spectrum}
Before estimating the anti-proton flux at the Earth, one needs to know the anti-proton spectrum before propagation, that is the one after the dark matter annihilation into scalar $h$ and pseudo-scalar $a$, their subsequent decay into quark pairs and their hadronization.
The first step is easy: it is two body annihilation so all the 4-momenta are set by kinematics. In the approximation that the dark matter particles annihilate at rest in the halo frame the energy of particle 1 (either $h$ or $a$), is
\begin{equation}
E_1 = \frac{4 m_\chi^2 + m_1^2 - m_2^2}{4 m_\chi} \nonumber
\end{equation}
and the norm of its momentum is :
\begin{equation}
k = \frac{\sqrt{\lambda\left(4 m_\chi^2 , m_1^2 , m_2^2\right)}}{4 m_\chi} \nonumber
\end{equation}
which are enough to change from the halo frame to the rest frame of particle 1: ($\gamma,\beta$)=($E_1/m_1,k/E_1$).
In the rest frame of particle 1, the quarks it decays into have energy $E_q^\ast = m_1/2$ and momentum $|k_q^\ast| = \sqrt{m_1^2/4 - m_q^2}$. One finally gets the energy of the quarks in the halo frame:
\begin{eqnarray}
E_q = \frac{E_1}{2} - \cos(\theta) \frac{\sqrt{\lambda\left(4 m_\chi^2 , m_1^2 , m_2^2\right)}}{8 m_\chi} \sqrt{1 - \frac{4 m_q^2}{m_1}} = \frac{E_1}{2} - \cos(\theta) \mathcal{E}.\nonumber
\end{eqnarray}
So the energies of the quarks and anti-quarks coming from the decay of particle 1 are evenly distributed between $\frac{E_1}{2} - \mathcal{E}$ and $\frac{E_1}{2} + \mathcal{E}$. Finally one gets  the probability of having an anti-proton of energy $E_{\overline{p}}$ from a quark of energy $E_q$ $f(E_q,E_{\overline{p}})$ thanks to the PYTHIA\footnote{For this work we made use of version 6.4.24 with CDF tune A} package~\cite{2006JHEP...05..026S}.
So finally the anti-proton spectrum after the dark matter annihilation is:
\begin{equation}
\mathcal{F}(E_{\overline{p}}) \!\!=\!\!\! \sum_{i=1,2}\!\! \sum_{q} BR_{i,q} \int_{\frac{E_i}{2} - \mathcal{E}}^{\frac{E_i}{2} + \mathcal{E}}\!\! \frac{f(E_q,E_{\overline{p}})+f(E_{\overline{q}},E_{\overline{p}})}{2 \mathcal{E}} dE_q, \nonumber
\end{equation}
where the first sum is done over the annihilation products (1,2)=($h,a$), and the second sum is over all the quark flavours for which $2 m_q \leqslant m_i$ is satisfied. The branching ratios $BR_{i,q}$ depend on the particle physics model considered.

\begin{figure}[!h]
\begin{center}
\includegraphics[width=0.48\textwidth]{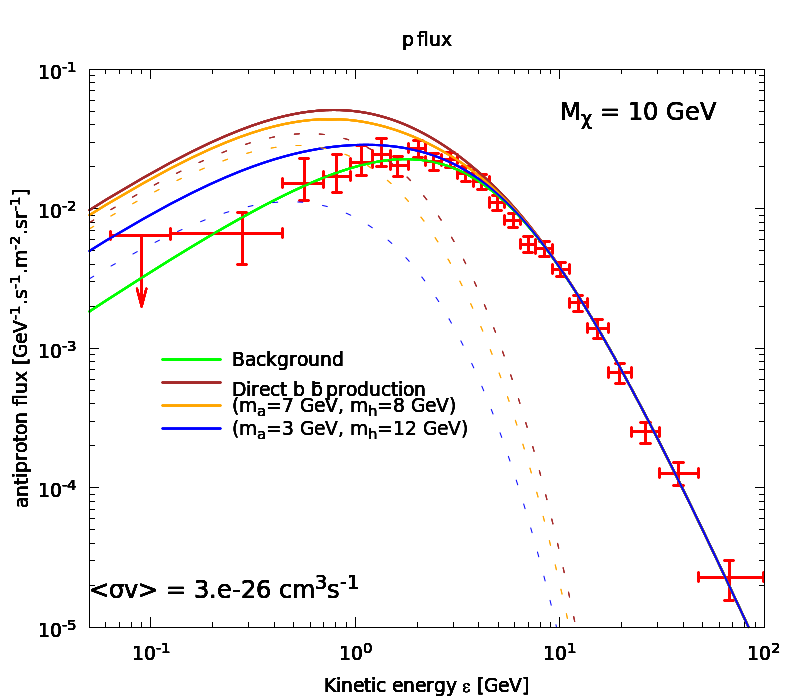}
\end{center}
\caption{Anti-proton flux at the Earth as a function of particle kinetic energy. Data are from PAMELA. Green line: background estimation. Brown, yellow and blue correspond to direct annihilation into $b\bar{b}$, annihilation into a scalar and a pseudo-scalar of masses of 7 and 8 GeV or 3 and 12 GeV, respectively. Dashed are signal only, lines are signal plus background.
\label{fig:flux}}
\end{figure}

Finally, one has to propagate the anti-proton from the annihilation place to the Earth. The model used here to describe this propagation has been detailed at length in many other papers~\cite{2006astro.ph..9522M,2006astro.ph.12714M,2009PhRvL.102g1301D}. In this model, the charged cosmic rays diffuse off the inhomogeneities of the Galactic magnetic field, they interact with the interstellar gas when they cross the disk and finally reach the Earth. This model has been shown to be extremely accurate in describing many cosmic ray species and, in particular, describes very well the anti-proton astrophysical background. This background is due to the spallation of cosmic ray protons and $\alpha$ on the interstellar hydrogen and helium (secondary cosmic rays). Moreover, when considering anti-protons it is important to also take into account tertiary cosmic rays from inelastic scattering of cosmic ray anti-protons. The prediction is in very good agreement with present data (as one can see from Figure~\ref{fig:flux}) and suffers very little from the uncertainties affecting the propagation parameters.

In order to determine whether or not a point in our parameter space is in agreement with anti-proton constraints, we simply summed the astrophysical prediction and the dark matter component and checked if the total was flux was higher than the PAMELA data.

\section{Results and conclusions}

\begin{figure}[!h]
\begin{center}
\includegraphics[width=0.48\textwidth]{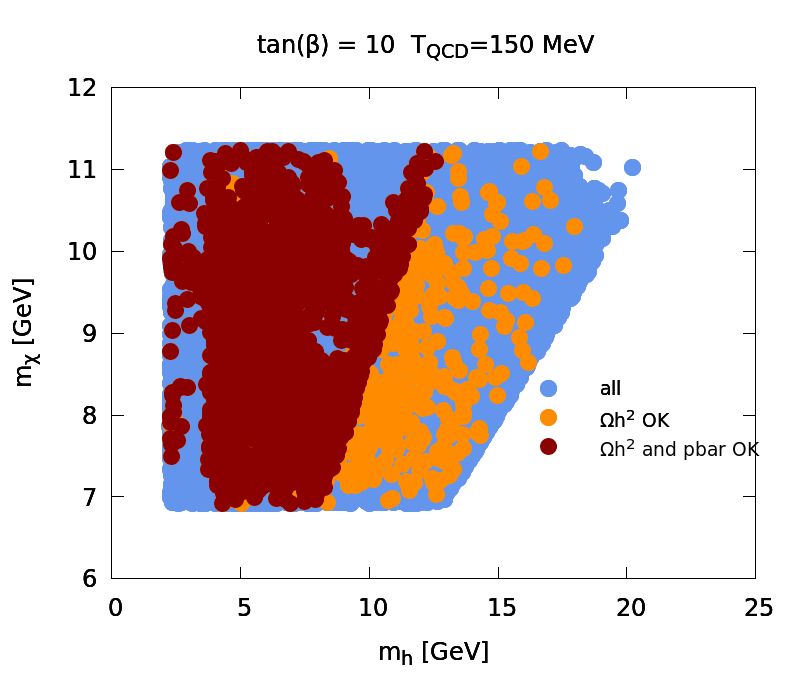}
\includegraphics[width=0.48\textwidth]{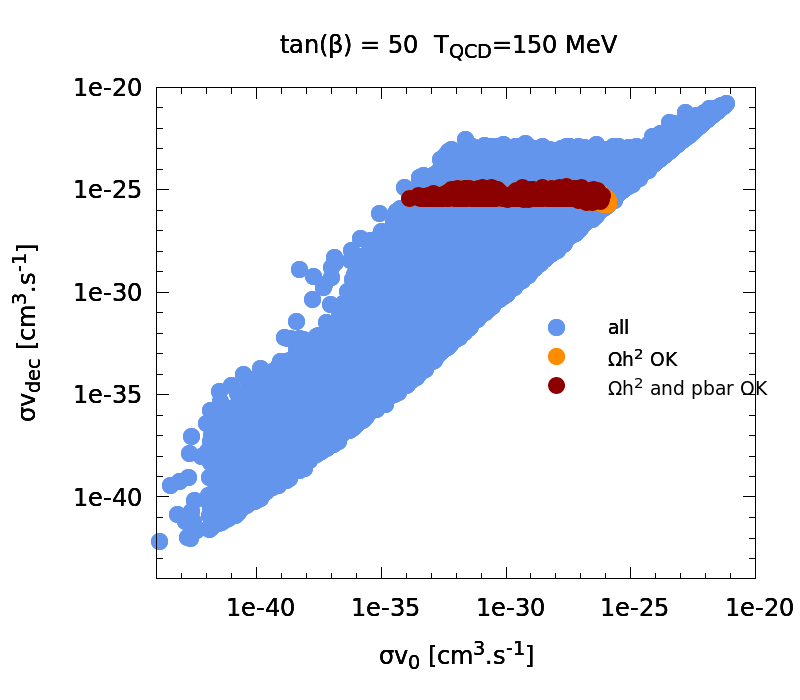}
\end{center}
\caption{Probing the parameter space. Left: scalar and singlino masses. Right : annihilation cross-section at time of decoupling and in the halo. Blue, points in the CoGeNT region, orange: correct relic density but excessive anti-proton production, brown, correct relic density and correct anti-proton flux.
\label{fig:exclus}}
\end{figure}

As one can see from Figure~\ref{fig:exclus}, the parameter space which gives a signal in the CoGeNT region, gives the correct relic density and in the same time does not give a too high cosmic ray anti-proton flux is quite small. This is not the place for a thorough study of the parameter space (see more details in~\cite{mainpaper}), however some features can be stressed thanks to Figure~\ref{fig:exclus}.

Large masses of the scalar particle $h$ are forbidden by cosmic ray constraints. Indeed, when $m_h$ is too large, the quark spectrum is not very much boosted (the quark pairs are almost produced at rest) so the anti-proton spectrum is peaked and easily exceed the observation, as if the dark matter particles were annihilating directly into quark pairs. Conversely, very low scalar masses cannot be constrained as they cannot decay into $b\bar{b}$ pairs. From the right panel of Figure~\ref{fig:exclus}, it appears that, quite naturally, high annihilation cross sections at zero velocity ($<\sigma v_0>~~\gtrsim~10^{-27}$~cm$^3$.s$^{-1}$) are excluded but not very low ones. The large discrepancy that can occur between annihilation cross sections at rest and at time of decoupling shows that the relic density can sometimes been set by the $t$-channel annihilation rather than by the $s$-channel, alleviating the constraints from anti-proton data.

Of course, one needs to repeat this work for the different possible dark matter halo profiles (here we made use of the profile proposed by \cite{2010JCAP...08..004C}) and for different propagation parameter sets. However one can already conclude that interpreting the recent direct detection experiments results in term of dark matter is quite challenging. The compatible parameter space, even when the dark matter annihilation does not directly goes into quark pairs, is extremely reduced. If the CoGeNT, DAMA/Libra result were to be confirmed by CDMS, Xenon and Edelweiss, it is interesting to stress that the absence of signal in the cosmic anti-proton channel is very enlightening from the point of view of the nature of dark matter as it would put very strict constraints on masses and couplings which could be challenging for LHC.

\section*{Acknowledgments}

This work was supported by the Spanish MICINN’s  Consolider-Ingenio 2010
Programme under grants CPAN CSD2007-00042 and MultiDark CSD2009-00064. 
We also acknowledge the support of the
MICINN under grant FPA2009-08958, the Community of Madrid under grant
HEPHACOS S2009/ESP-1473, and the European Union under the Marie Curie-ITN
program PITN-GA-2009-237920.

\bibliography{bibli}

\end{document}